\documentclass[aip,jcp,amsmath,amssymb,reprint,floatfix]{revtex4-1}

\usepackage{graphics}
\usepackage{graphicx}
\usepackage{color}
\usepackage{amsmath}
\usepackage{subfigure}
\usepackage[utf8]{inputenc}
\usepackage{siunitx}

\begin{document}

\title{The free-energy landscape of a mechanically bistable DNA origami}
\author{Chak Kui Wong}
\affiliation{Physical and Theoretical Chemistry Laboratory, Department of Chemistry, University of Oxford, South Parks Road, Oxford, OX1 3QZ, United Kingdom}
\author{Jonathan P. K. Doye}
\affiliation{Physical and Theoretical Chemistry Laboratory, Department of Chemistry, University of Oxford, South Parks Road, Oxford, OX1 3QZ, United Kingdom}

\date{\today}

\begin{abstract}
Molecular simulations using coarse-grained models allow the structure, dynamics and mechanics of DNA origamis to be comprehensively characterized. Here, we focus on the free-energy landscape of a jointed DNA origami that has been designed to exhibit two mechanically stable states and for which a bistable landscape has been inferred from ensembles of structures visualized by electron microscopy. Surprisingly, simulations using the oxDNA model predict that the defect-free origami has a single free-energy minimum. The expected second state is not stable because the hinge joints do not simply allow free angular motion but instead lead to increasing free-energetic penalties as the joint angles relevant to the second state are approached. This raises interesting questions about the cause of this difference between simulations and experiment, such as how assembly defects might affect the ensemble of structures observed experimentally.
\end{abstract} 

\maketitle


\section{Introduction}

The DNA origami technique \cite{Rothemund06,Douglas09,Dey21} has emerged as a powerful approach to create self-assembled nanostructures and nanodevices. One area of increasing interest is their use as nanoscale mechanical devices \cite{DeLuca20}. In this field of DNA ``mechanotechnology'' \cite{Blanchard19} DNA constructs are used to generate \cite{Nickels16b,Le16,Su21}, transmit \cite{Pfitzner13} and sense \cite{Dutta18} nanoscale forces. One approach is through the use of flexible DNA origami that can undergo large-scale structural changes \cite{Zhou14,Marras15,Zhou18}; for example, an externally-actuated shape change could be used to apply a force \cite{Wang21b}, or a force could be sensed through the change in structure it induces \cite{Funke16c,Hudoba17,Ke16}. 

For such applications it is important to fully understand the shape changes that origamis can undergo and how the thermodynamics and mechanics of these shape changes can be controlled through the origami design. In this regards, computational approaches have the potential to play an important role as they can provide direct insight into how the microscopic features of the design impact the overall properties of the origami \cite{Shi17,Sharma17,Shi19}.

Here, we explore these issues further by providing a detailed simulation analysis of the jointed DNA origami introduced by Zhou {\it et al.} \cite{Zhou15} that is designed to be bistable with the free-energy barrier between the two stable states being a result of the internal mechanical stresses associated with intermediate states. Observations of ensembles of origami structures by electron microscopy seem to have confirmed the essential bistability of the design. In particular, we will compute the free-energy landscape of this system and explore the role of the properties of the individual joints in determining the overall behaviour. The study will raise interesting questions about the potential role of defects on the properties of these origami systems.

\section{Materials and Methods}

To model the DNA nanostructures we use the oxDNA coarse-grained model \cite{Ouldridge11,Sulc12,Snodin15}. In particular, we use the second version of the model that has been fine-tuned to improve the modelling of origami structure \cite{Snodin15}. As a results of its accurate description of the structural \cite{Snodin19} and mechanical \cite{Chhabra20} properties of DNA origami, it has been widely used in the field of DNA nanotechnology, particularly to describe origamis with increasingly functionally complex properties that are otherwise hard to predict \cite{Benson18,Berengut19,Berengut20,Engel20, Yao20,Huang21,Li21,Su21}. It is particularly well-suited to study jointed origami, as it can also capture the important local features of the junction; e.g.\ an accurate description of the mechanical properties of single-stranded DNA \cite{Sulc12} is important to model the role of the single-stranded linkers on joint flexibility. Furthermore, oxDNA is computationally efficient enough to be combined with techniques such as umbrella sampling \cite{Wong22} and metadynamics \cite{Kaufhold22} to compute the free-energy landscapes associated with mechanical deformations of DNA origamis.

In the oxDNA model, each nucleotide is represented by a rigid body. The nucleotides interact through a pairwise potential which has terms representing backbone connectivity, excluded-volume interactions, hydrogen bonding between base pairs, stacking interactions, and electrostatic interactions between backbone phosphate groups. Solvent is modelled implicitly as a dielectric continuum, and we choose to use a salt concentration of [Na$^+$]\,=\,1.0\,M, which is representative of the high salt conditions typically used for DNA nanotechnology.

\begin{figure*}
    \centering
    \includegraphics[width=6.7in]{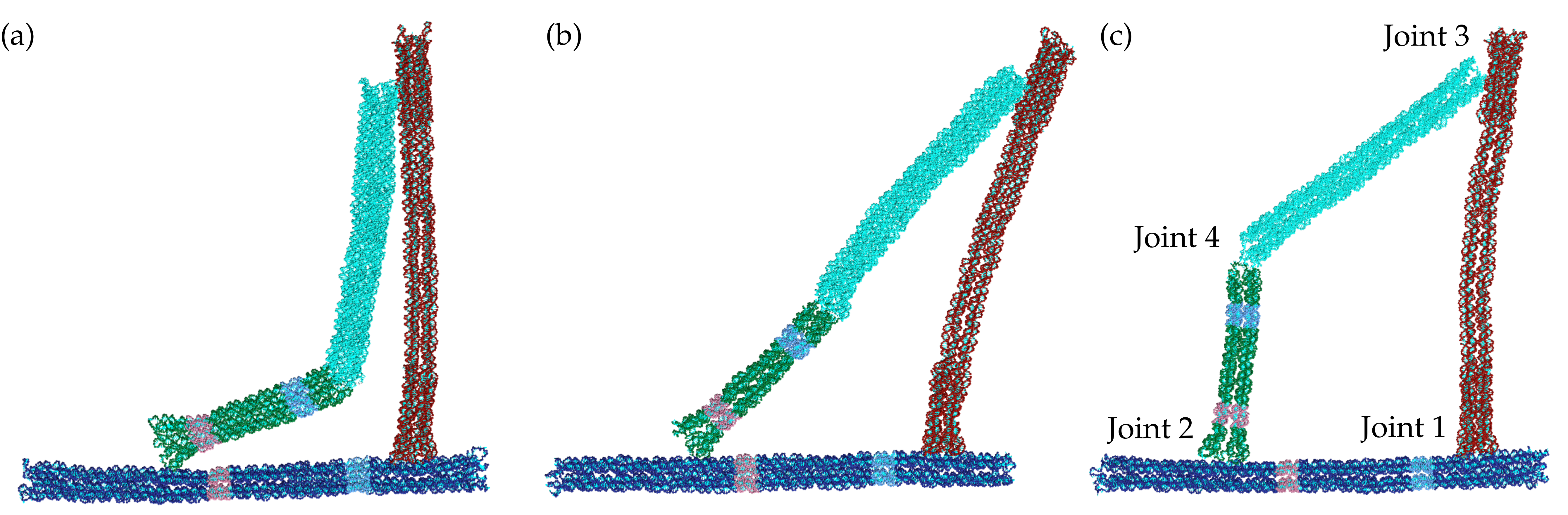}
    \caption{Representative oxDNA configurations of the origami in \textbf{(a)} the closed state S1 ($\theta \approx 20^{\circ}$) \textbf{(b)} the intermediate state U ($\theta \approx 45^{\circ}$) and \textbf{(c)} the open state S2 ($\theta \approx 80^{\circ}$). The four joints are numbered in (c).}
    \label{fig:castro_configs2}
\end{figure*}

The oxDNA simulation code was used to perform molecular dynamics simulations in the canonical ensemble, in particular making use of its GPU implementation \cite{Rovigatti15}. To sample the free-energy landscapes we used umbrella sampling in a similar manner to Ref.\ \cite{Wong22}. For each free-energy landscape we ran a series of between 20 and 80 simulations each of which sampled a separate order-parameter window due to the placement of a harmonic biasing potential in the order parameter at the centre of the window. The free-energy landscapes were constructed from the order-parameter probability distributions for each window using the weighted-histogram analysis method (WHAM) \cite{Kumar92}. We iterated the sets of umbrella sampling simulations until the WHAM-generated landscapes no longer changed significantly.
Additional simulation and analysis details can be found in the Supplementary Material.


\section{Results}

\subsection{Design}
The bistable jointed origami consists of four rod-like blocks of parallel double helices based on a hexagonal lattice that are linked through scaffold connections forming a roughly planar arrangement (Figure \ref{fig:castro_configs2}). In Ref.\ \cite{Zhou15} these blocks have been termed the frame (dark blue), compliant (red), crank (green) and coupler (cyan) blocks. The central section of the compliant block is a 6-helix bundle and is designed to bend more easily than the other blocks which are all based on 10-helix bundles. For example, using the oxDNA model such a 10-helix bundle has been found to have a bending persistence length that is about three times larger than a 6-helix bundle \cite{Chhabra20}.

Joint 1 between the frame and compliant block is designed to be rigid, whereas joints 2--4 are designed to behave like hinges allowing free rotational motion of the blocks in the plane of the structure. Joints 2 and 3  are connected by short single-stranded linkers on one side of the block, so that the block can rotate about that edge. The linkers for joint 4 are somewhat longer to provide additional flexibility so that the joint can bend both inwards and outwards.

The origami has been designed to have a free-energy landscape with two minima (termed the ``closed'' state S1 and the ``open'' state S2) as a function of $\theta$ the angle between the crank and the frame. Since joint 1 is rigid, in order for the origami to transition between the two stable states, the compliant block needs to bend outwards to accommodate the extra length between the ends of the crank and the coupler blocks. When the crank and the coupler blocks are collinear, the bending in the compliant block is at its maximum, and so this intermediate state U is expected to be unstable, i.e.\ a local maximum on the free-energy landscape. A simple mechanical model that accounts for the cost of bending the compliant block, while assuming free rotation at joints 2--4 predicts a free-energy barrier of 3--4\,$k_B T$ \cite{Zhou15}. Representative oxDNA configurations at the $\theta$ values corresponding to S1, U and S2 are depicted in Fig.\ \ref{fig:castro_configs2}. 

Experimentally, the free-energy landscape for this origami has been inferred from ensembles of structures visualized using electron microscopy \cite{Zhou15}. In particular, the probability distribution for $\theta$ was measured (the free-energy landscape is proportional to the log of this distribution). The main peaks at 28$^\circ$ and 80$^\circ$ correspond to states S1 and S2, but surprisingly there was a small sub-peak in the probability distribution at intermediate values of $\theta$ rather than a minimum. It was suggested that this sub-population in the U state might be a result of coaxial stacking between the ends of the crank and coupler blocks when they are collinear. 

\subsection{Free-Energy Landscape}
\begin{figure*}
    \centering
    \includegraphics[width=5.4in]{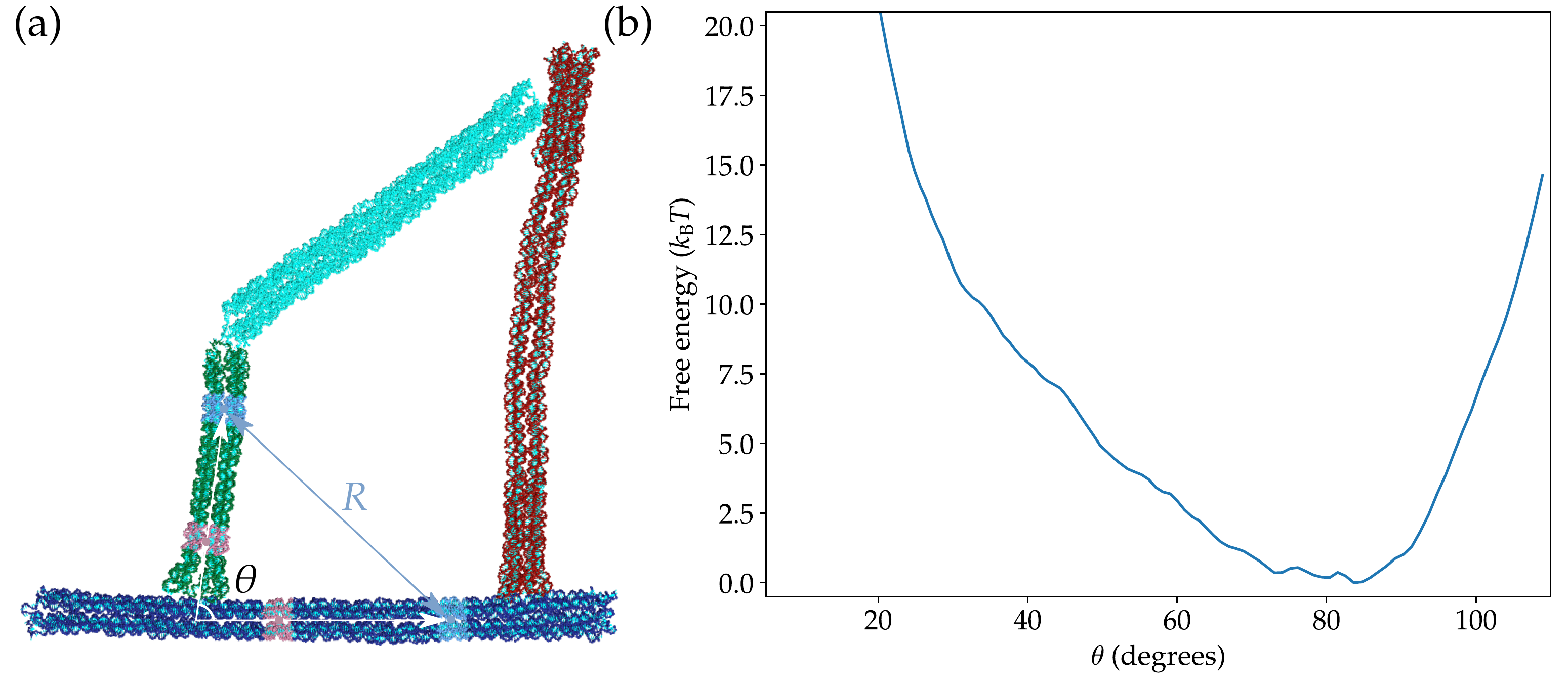}
    \caption{\textbf{(a)} A configuration of the origami illustrating the definition of the order parameter $R$ as the distance between the centres of mass of the two groups of nucleotides coloured in light blue. The value of $\theta$ was tracked in the simulations, and is defined as the angle between the two vectors coloured in white. The vectors pass through the centres of mass of the two groups of nucleotides coloured in light red and light blue in the crank block and the frame block, respectively. \textbf{(b)} The free-energy landscape of the origami as a function of $\theta$.}
    \label{fig:castro_FEL_and_order_param}
\end{figure*}

The free-energy landscape was calculated using umbrella sampling and WHAM. We chose the order parameter as the distance $R$ between the centres of mass of two groups of nucleotides in the crank block and the frame block, respectively. The groups of nucleotides are highlighted in Figure~\ref{fig:castro_FEL_and_order_param}a. 
We chose to use $R$ as the order parameter rather than $\theta$ because calculating the forces on the nucleotides due to the umbrella potential is then much simpler. The values of $R$ and $\theta$ were tracked during the umbrella sampling simulations so that we can then transform the free-energy landscape from a function of $R$ to a function of $\theta$. The resulting landscape is shown in Figure~\ref{fig:castro_FEL_and_order_param}b. 

Contrary to previous expectations, the landscape obtained using the oxDNA model only shows a single minimum at $\theta \approx 80^{\circ}$, which corresponds to state S2. There is no sign of any second minimum corresponding to state S1. Instead, the free energy continuously increases as $\theta$ decreases. At $\theta \approx 20^{\circ}$, where state S1 is expected to occur, the free energy is more than 20\,$k_B T$. This is significantly different from the expectation based on the experimental results, where about 30\% of the origamis were observed to be in the closed state.

\subsection{Free-energy Decomposition}
\begin{figure*}
    \centering
    \includegraphics[width=5.4in]{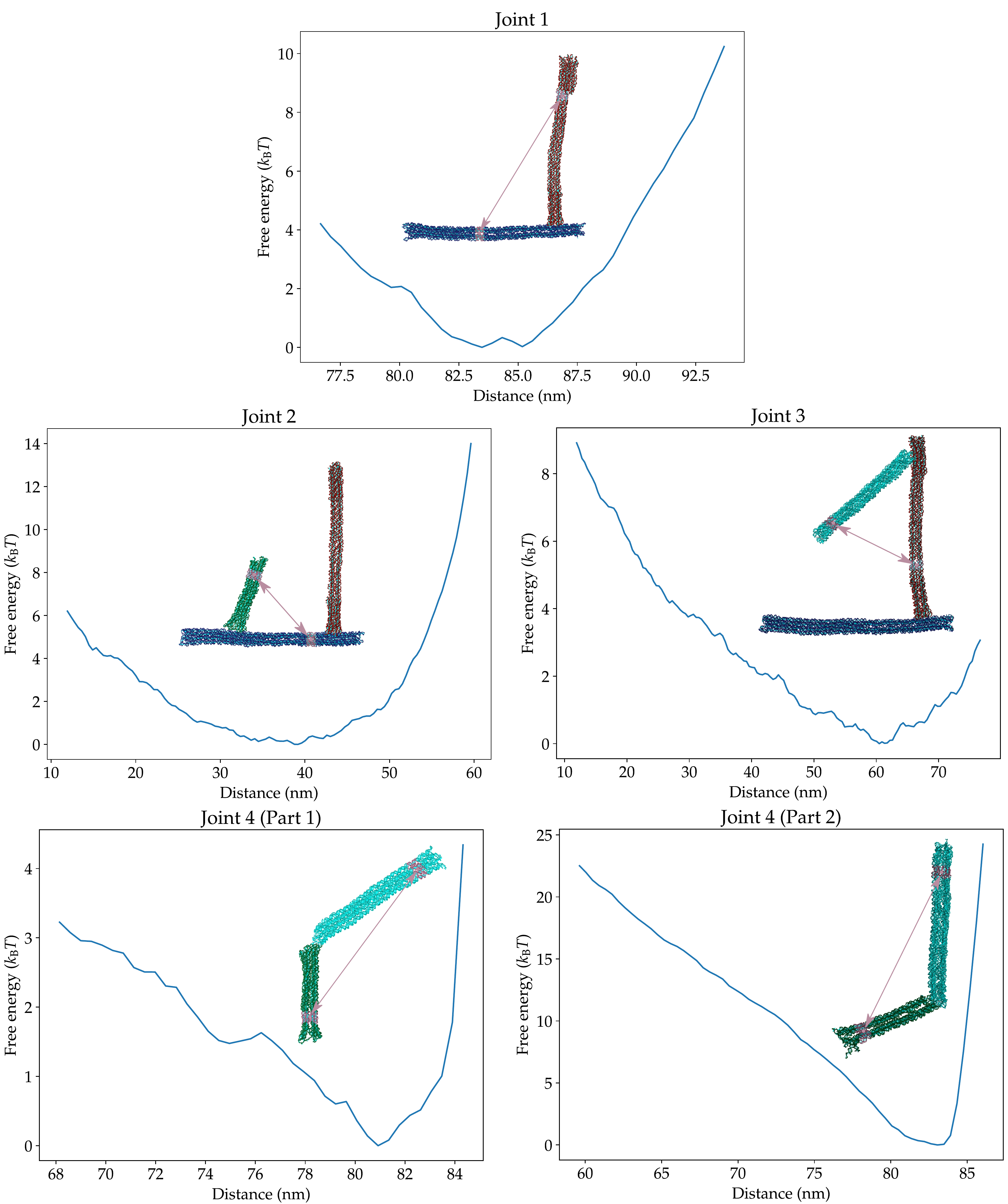}
    \caption{The free-energy landscapes associated with each individual joint. The configurations illustrate the sub-systems used to compute these landscapes and the definitions of the distance order parameters. The values of these order parameters in the depicted configurations are 86\,nm, 45\,nm, 51\,nm, 78\,nm and 70\,nm, respectively. The landscape associated with joint 4 has been
    computed in two parts corresponding to whether the joint is pointing outwards (1) or inwards (2). Only when joint 4 is straight do these sub-landscapes sample equivalent configurations.}
    \label{fig:castro_component_all}
\end{figure*}

To better understand the form of the oxDNA free-energy landscape, and in particular why there is no minimum corresponding to state S1, we attempted to compute the contributions to the free-energy landscape from the four joints. In the complete origami the motion of the four joints is coupled due to overall structure of the origami. Therefore, to isolate the behaviour of the individual joints we performed a series of simulations with one or more blocks of the origami removed (see the configurations in Fig.\ \ref{fig:castro_component_all}). Similar to the previous section, we then defined distance order parameters between the centres of mass of groups of nucleotides to probe the motion of a given joint.
The definitions of the order parameters and their respective free-energy landscapes are shown in Figure~\ref{fig:castro_component_all}.

\begin{figure}
    \centering
    \includegraphics[width=3.3in]{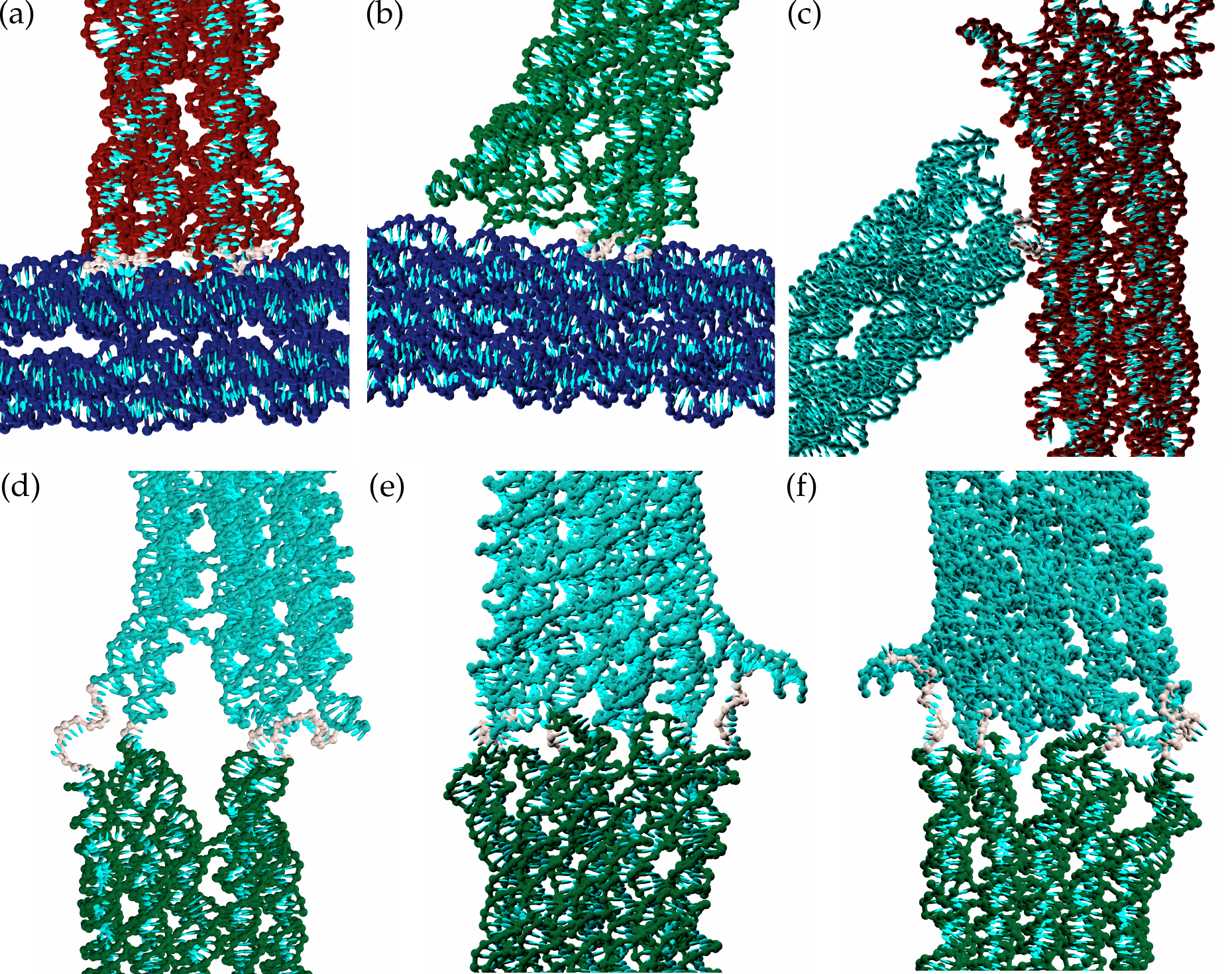}
    \caption{Close-ups of each joint for the configurations illustrated in Fig.\ \ref{fig:castro_component_all}. Specifically, these are for \textbf{(a)} Joint 1 \textbf{(b)} Joint 2 \textbf{(c)} Joint 3 \textbf{(d)} Joint 4 (part 1) \textbf{(e)} Joint 4 (part 2, view from outside) \textbf{(f)} Joint 4 (part 2, view from inside).
    (The outside and inside of joint 4 are defined relative to the quadrilateral formed by the four blocks.) Single-stranded sections of the scaffold that link the blocks are coloured grey. (d) and (e)/(f) clearly show the difference in steric repulsion between the helices near the joint for open and closed configurations of joint 4.}
    \label{fig:castro_closeup}
\end{figure}

Joint 1 is designed to be rigid and, as expected, the free-energy minimum is when the frame and compliant blocks are perpendicular and unbent. The free-energy cost of changing the distance order parameter comes mainly from the bending of the compliant block instead of a change in the angle at the joint. This degree of freedom is the intended mode of stress accumulation in the original design that leads to disfavouring of the intermediate state.

Joints 2 and 3 are both designed to act as hinges and have a similar local structure. However, as their free-energy landscapes show, the hinges are not completely flexible. They have preferred opening angles at around $55^{\circ}$ and $70^{\circ}$, respectively, and there is a non-negligible free-energy cost when the angles deviate from the preferred values. In particular, there are significant steric repulsions between the helices in the two blocks when the opening angle goes beyond $90^\circ$ or approaches zero. 

Joint 4 was designed to be more flexible as the joint needs to bend both inward and outward to allow both the open and closed states to be easily accessed. The landscape for the joint has been computed in two sections because the distance order parameter on its own cannot distinguish whether the interior angle of the joint is obtuse (as in the open state) or reflex (as in the closed state). 
We find that the joint prefers to be roughly straight with an increasing free-energy penalty for more bent configurations. (Note the sharp rise in the free energy at large values of the distance order parameter is due to the stretching of the straight joint and is not relevant to the behaviour of the complete origami.) Particularly noteworthy is the asymmetry between bending the junction outwards and inwards, with a much higher cost for the inward bending due to the resulting steric repulsions between the helices near to the joint (Fig.\ \ref{fig:castro_closeup}).

\begin{figure}
    \centering
    \includegraphics[width=3.3in]{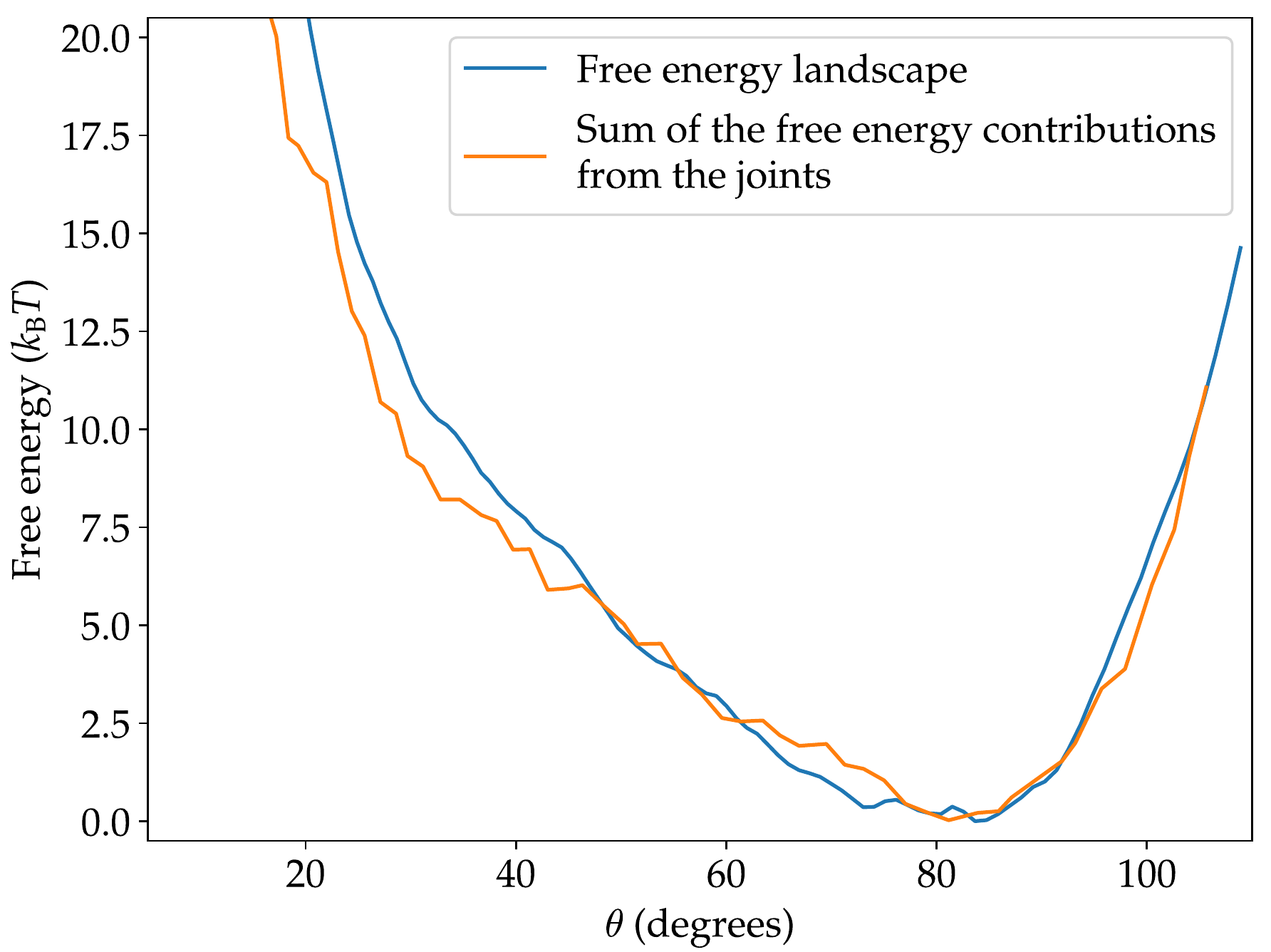}
    \caption{Comparison of the free-energy landscape and the sum of the contributions from the joints.}
    \label{fig:castro_FEL_comparison}
\end{figure}

By keeping track of the distance order parameters of each joint as well as $R$ and $\theta$ in simulations of the complete origami, we can transform the joint free-energy landscapes to be functions of $\theta$ (Fig.\ S1). We next check if the free energy of the complete origami where the motion of the joints is coupled can be represented as the sum of the contributions from the individual joints. The comparison in Fig.\ \ref{fig:castro_FEL_comparison} shows that this is a very good approximation. 
Having established this, we now use these landscapes to understand why oxDNA predicts the closed state S1 to be unstable.

\begin{figure}
    \centering
    \includegraphics[width=3.3in]{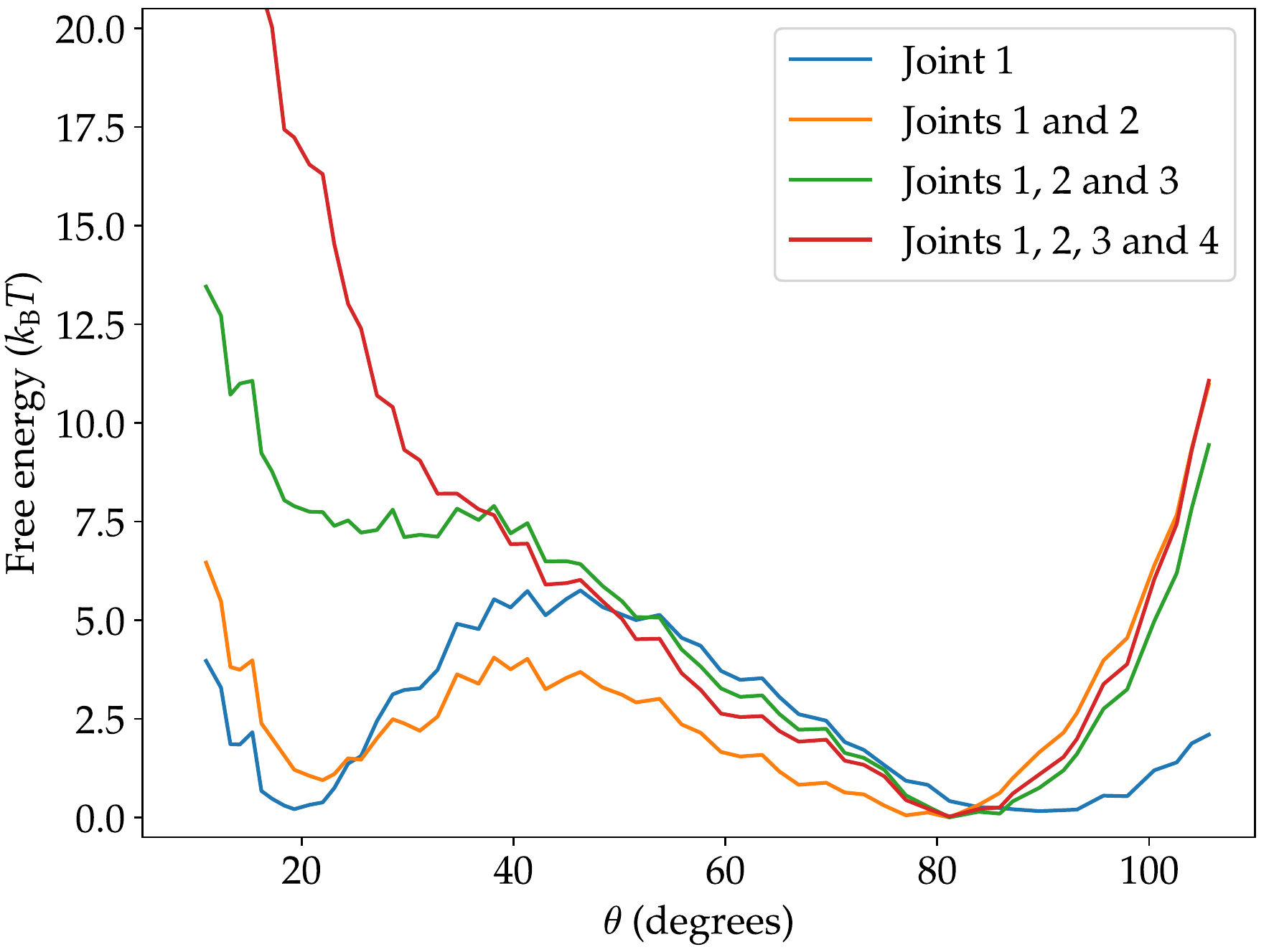}
    \caption{The Free-energy contributions of each joint to the overall free-energy landscape.}
    \label{fig:castro_FEL_decomposed}
\end{figure}

The contribution to the free energy of the complete origami associated with joint 1 shows two minima at $\theta \approx 20^{\circ}$ and at $\theta \approx 90^{\circ}$, and a maximum at around $\theta \approx 45^{\circ}$ (Fig.\ \ref{fig:castro_FEL_decomposed}). These positions generally agree with the angles for the S1, U and S2 states observed experimentally \cite{Zhou15}. Furthermore, this component of the free-energy landscape is very similar to that predicted by the mechanical model used in that paper, albeit with a slightly larger barrier that is probably due to the somewhat high values of persistence lengths that oxDNA predicts for DNA origami \cite{Chhabra20}. These results indicate that the designed mechanism for stress accumulation due to bending of the compliant block can be captured by oxDNA.

The free-energy contribution of joint 2 is a broad minimum centred around the preferred opening angle of the joint (Fig.\ S1), and so adding this contribution does not significantly alter the general shape of the landscape. It slightly lowers the barrier and penalizes configurations with more extreme values of $\theta$, particularly those with $\theta > 90^{\circ}$, because
of the steric repulsions between the crank and frame blocks at these joint angles.

Further including the contribution from joint 3 results in an appreciable change in the shape of the landscape. Most noticeably, the minimum corresponding to S1 becomes a flat region with a higher free energy. Its contribution has a more significant effect than joint 2 because the structure of the origami necessitates smaller opening angles at joint 3 than joint 2. It is these small opening angles, and their associated free-energy cost, that particularly lead to the destabilization of S1.

Finally, including the contribution from joint 4 completely removes the second minimum,
because the inward bending of this joint that is associated with the closed configuration leads to a significantly larger free-energy cost.

\section{Discussion}


In this paper we have used the oxDNA coarse-grained model to compute the free-energy landscape of a jointed multi-block DNA origami that was designed to have two stable states. Somewhat surprisingly, the oxDNA landscape shows a single minimum corresponding to the open configuration. These results present a substantially different picture than from experiment for which a significant population of origami in the closed state was observed. 

Although the bending of the compliant block associated with intermediate configurations, the designed stress accumulation mechanism, leads to a bistable landscape associated with joint 1, that the other three joints do not behave like completely free hinges in oxDNA leads to the destabilization of the closed state with the small opening angles at joint 3 and the inward bending of joint 4 being particularly unfavourable. The increase in the free energy of the hinge joints as their joint angles move away from their preferred values is due both to direct steric repulsions and the constraints increasingly placed on the configuration of the helices and linkers near the joint. We note that these effects are consistent with previous oxDNA results on jointed DNA nanostructures \cite{Sharma17, Shi19} that have also found narrower distributions for the flexible components than observed in electron microscopy images; i.e.\ joints in oxDNA seem to be less flexible than inferred from experiment.

So what are the possible causes of the seeming disagreement between the simulations and experiment. One possibility is simply that the oxDNA model is deficient in its description of these structures. However, the behaviour of the joints must be a reflection of the structural and mechanical properties of the helices and single-stranded DNA at the joints, which are all things that oxDNA describes well. Moreover, these are not small effects; oxDNA is predicting that the closed state of the origami is $20\,k_B T$ higher in energy than the open state.

Another possible source is the heterogeneity in the ensemble of assembled origami in experiment. Whereas the simulations model a defect-free origami, there are likely to be a range of assembly and other defects in the experimental origami. These assembly defects probably not only include missing staples but also defects in the routing of the strands. The latter may be particularly relevant to multi-block origami, such as those studied here, which possess many scaffold links between the blocks. For example, if the blocks nucleate and assemble independently \cite{Marras16}, then if the assembly of one block happens to lead to the topological entanglement of the external scaffold loops, such a defect is likely to be locked in. Furthermore, the assembly of further blocks might push these defects to the joints, potentially substantially changing the properties of the joint. Although there has been some quantification of missing staples \cite{Wagenbauer14,Myhrvold17,Strauss18}, there is little information about other possible defects.
Thus, it is conceivable that the population of closed configuration corresponds to origami that are locked into this structure due to assembly defects or that have more flexible joints due to missing staples. 

There is some evidence of this heterogeneity in the experimental results. For example, it is not clear why there was a small sub-population of U states even though such structures were designed to be at a maximum on the free-energy landscape. Again, it is possible that they might be locked in to these configurations by defects. 
Furthermore, that a fraction of the open configurations do not close on the addition of actuating strands that lead to binding between the frame and crank blocks, and the coupler and compliant blocks again suggests the presence of certain defects preventing the closing. 

One possible way of to help address these issues and to quantify the heterogeneity in such ensembles of flexible origamis might be to use single-molecule techniques that can follow the large-scale dynamical motion of individual origamis. One could also use simulations to explore the effects of different types of defects on origami properties.






\begin{acknowledgments}
This research was funded by the Croucher Foundation.
The authors acknowledge the use of the University of Oxford Advanced Research Computing (ARC) facility (http://dx.doi.org/10.5281/zenodo.22558).
\end{acknowledgments}






\begin{thebibliography}{999}

\bibitem[Rothemund(2006)]{Rothemund06}
Rothemund, P.W.K.
\newblock Folding {DNA} to create nanoscale shapes and patterns.
\newblock {\em Nature} {\bf 2006}, {\em 440},~297--302.

\bibitem[Douglas \em{et~al.}(2009)Douglas, Dietz, Liedl, H\"{o}gberg, Graf, and
  Shih]{Douglas09}
Douglas, S.M.; Dietz, H.; Liedl, T.; H\"{o}gberg, B.; Graf, F.; Shih, W.M.
\newblock Self-assembly of {DNA} into nanoscale three-dimensional shapes.
\newblock {\em Nature} {\bf 2009}, {\em 459},~414--418.

\bibitem[Dey \em{et~al.}(2021)Dey, Fan, Gothelf, Li, Lin, Liu, Nijenhuis,
  Sacca, Simmel, Yan, and Zhan]{Dey21}
Dey, S.; Fan, C.; Gothelf, K.V.; Li, J.; Lin, C. anc~Liu, L.; Liu, N.;
  Nijenhuis, M.A.D.; Sacca, B.; Simmel, F.C.; Yan, H.; Zhan, P.
\newblock DNA origami.
\newblock {\em Nat. Rev. Methods Primers} {\bf 2021}, {\em 1},~13.

\bibitem[DeLuca \em{et~al.}(2020)DeLuca, Shi, Castro, and Arya]{DeLuca20}
DeLuca, M.; Shi, Z.; Castro, C.E.; Arya, G.
\newblock Dynamic {DNA} nanotechnology: Toward functional nanoscale devices.
\newblock {\em Nanoscale Horiz.} {\bf 2020}, {\em 5},~182--201.

\bibitem[Blanchard and Salaita(2019)]{Blanchard19}
Blanchard, A.T.; Salaita, K.
\newblock Emerging uses of {DNA} mechanical devices.
\newblock {\em Science} {\bf 2019}, {\em 365},~1080--1081.

\bibitem[Nickels \em{et~al.}(2016)Nickels, W\"{u}nsch, Holzmeister, Bae, Kneer,
  Grohmann, Tinnefeld, and Liedl]{Nickels16b}
Nickels, P.C.; W\"{u}nsch, B.; Holzmeister, P.; Bae, W.; Kneer, L.M.; Grohmann,
  D.; Tinnefeld, P.; Liedl, T.
\newblock Molecular force spectroscopy with a {DNA} origami-based nanoscopic
  force clamp.
\newblock {\em Science} {\bf 2016}, {\em 354},~305--307.

\bibitem[Le \em{et~al.}(2016)Le, Luo, Darcy, Lucas, Goodwin, Poirier, and
  Castro]{Le16}
Le, J.V.; Luo, Y.; Darcy, M.A.; Lucas, C.R.; Goodwin, M.F.; Poirier, M.G.;
  Castro, C.E.
\newblock Probing nucleosome stability with a {DNA} origami nanocaliper.
\newblock {\em ACS Nano} {\bf 2016}, {\em 10},~7073--7084.

\bibitem[Su \em{et~al.}(2021)Su, Brockman, Duan, Sen, Chhabra, Bazrafshan,
  Blanchard, Meyer, Andrews, Doye, Ke, Dyer, and Salaita]{Su21}
Su, H.; Brockman, J.M.; Duan, Y.; Sen, N.; Chhabra, H.; Bazrafshan, A.;
  Blanchard, A.T.; Meyer, T.; Andrews, B.; Doye, J.P.K.; Ke, Y.; Dyer, R.B.;
  Salaita, K.
\newblock Massively parallelized molecular force manipulation with on demand
  thermal and optical control.
\newblock {\em J. Am. Chem. Soc.} {\bf 2021}, {\em 43},~19466--19473.

\bibitem[Pfitzner \em{et~al.}(2013)Pfitzner, Wachauf, Kilchherr, Pelz, Shih,
  Rief, and Dietz]{Pfitzner13}
Pfitzner, E.; Wachauf, C.; Kilchherr, F.; Pelz, B.; Shih, W.M.; Rief, M.;
  Dietz, H.
\newblock Rigid {DNA} beams for high-resolution single-molecule mechanics.
\newblock {\em Angew. Chem. Int. Ed.} {\bf 2013}, {\em 52},~7766--7771.

\bibitem[Dutta \em{et~al.}(2018)Dutta, Zhang, Blanchard, Ge, Rushdi, Weiss,
  Zhu, Ke, and Salaita]{Dutta18}
Dutta, P.K.; Zhang, Y.; Blanchard, A.T.; Ge, C.; Rushdi, M.; Weiss, K.; Zhu,
  C.; Ke, Y.; Salaita, K.
\newblock Programmable multi-valent {DNA}-origami tension probes for reporting
  cellular traction forces.
\newblock {\em Nano Lett.} {\bf 2018}, {\em 18},~4803--4811.

\bibitem[Zhou \em{et~al.}(2014)Zhou, Marras, and Castro]{Zhou14}
Zhou, L.; Marras, A.E.; Castro, C.E.
\newblock Origami Compliant Nanostructures with Tunable Mechanical Properties.
\newblock {\em ACS Nano} {\bf 2014}, {\em 8},~27--34.

\bibitem[Marras \em{et~al.}(2015)Marras, Zhou, Su, and Castro]{Marras15}
Marras, A.E.; Zhou, L.; Su, H.J.; Castro, C.E.
\newblock Programmable Motion of DNA Origami Mechanisms.
\newblock {\em Proc. Natl. Acad. Sci. U. S. A.} {\bf 2015}, {\em
  112},~713--718.

\bibitem[Zhou \em{et~al.}(2018)Zhou, Marras, Huang, Castro, and Su]{Zhou18}
Zhou, L.; Marras, A.E.; Huang, C.M.; Castro, C.E.; Su, H.J.
\newblock Paper origami-inspired design and actuation of {DNA} nanomachines
  with complex motions.
\newblock {\em Small} {\bf 2018}, {\em 14},~1802580.

\bibitem[Wang \em{et~al.}(2021)Wang, Le, Crocker, Darcy, Halley, Zhao,
  Andrioff, Croy, Poirier, Bundschuh, and Castro]{Wang21b}
Wang, Y.; Le, J.V.; Crocker, K.; Darcy, M.A.; Halley, P.D.; Zhao, D.; Andrioff,
  N.; Croy, C.; Poirier, M.G.; Bundschuh, R.; Castro, C.
\newblock A nanoscale {DNA} force spectrometer capable of applying tension and
  compression on biomolecules.
\newblock {\em Nucleic Acids Res.} {\bf 2021}, {\em 49},~8987--8999.

\bibitem[Funke \em{et~al.}(2016)Funke, Ketterer, Lieleg, Schunter, Korber, and
  Dietz]{Funke16c}
Funke, J.J.; Ketterer, P.; Lieleg, C.; Schunter, S.; Korber, P.; Dietz, H.
\newblock Uncovering the forces between nucleosomes using {DNA} origami.
\newblock {\em Sci. Adv.} {\bf 2016}, {\em 2},~e1600974.

\bibitem[Hudoba \em{et~al.}(2017)Hudoba, Luo, Zacharias, Poirier, and
  Castro]{Hudoba17}
Hudoba, M.W.; Luo, Y.; Zacharias, A.; Poirier, M.G.; Castro, C.E.
\newblock Dynamic {DNA} origami device for measuring compressive depletion
  forces.
\newblock {\em ACS Nano} {\bf 2017}, {\em 11},~6566--6573.

\bibitem[Ke \em{et~al.}(2016)Ke, Meyer, Shih, and Bellot]{Ke16}
Ke, Y.; Meyer, T.; Shih, W.M.; Bellot, G.
\newblock Regulation at a distance of biomolecular interactions using a {DNA}
  nanoactuator.
\newblock {\em Nat. Commun.} {\bf 2016}, {\em 7},~10935.

\bibitem[Shi \em{et~al.}(2017)Shi, Castro, and Arya]{Shi17}
Shi, Z.; Castro, C.E.; Arya, G.
\newblock Conformational dynamics of mechanically compliant {DNA}
  nanostructures from coarse-grained molecular dynamics simulations.
\newblock {\em ACS Nano} {\bf 2017}, {\em 11},~4617--4630.

\bibitem[Sharma \em{et~al.}(2017)Sharma, Schreck, Romano, Louis, and
  Doye]{Sharma17}
Sharma, R.; Schreck, J.S.; Romano, F.; Louis, A.A.; Doye, J.P.K.
\newblock Characterizing the Motion of Jointed {DNA} Nanostructures Using a
  Coarse-Grained Model.
\newblock {\em {ACS} Nano} {\bf 2017}, {\em 11},~12426--12435.

\bibitem[Shi and Arya(2019)]{Shi19}
Shi, Z.; Arya, G.
\newblock {Free Energy Landscape of Salt-Actuated Reconfigurable DNA
  Nanodevices}.
\newblock {\em Nucleic Acids Res.} {\bf 2019}, {\em 48},~548--560.

\bibitem[Zhou \em{et~al.}(2015)Zhou, Marras, Su, and Castro]{Zhou15}
Zhou, L.; Marras, A.E.; Su, H.J.; Castro, C.E.
\newblock Direct Design of an Energy Landscape with Bistable DNA Origami
  Mechanisms.
\newblock {\em Nano Lett.} {\bf 2015}, {\em 15},~1815--1821.

\bibitem[Ouldridge \em{et~al.}(2011)Ouldridge, Louis, and Doye]{Ouldridge11}
Ouldridge, T.E.; Louis, A.A.; Doye, J.P.K.
\newblock Structural, Mechanical, and Thermodynamic Properties of a
  Coarse-Grained {DNA} Model.
\newblock {\em J. Chem. Phys.} {\bf 2011}, {\em 134},~085101.

\bibitem[\v{S}ulc \em{et~al.}(2012)\v{S}ulc, Romano, Ouldridge, Rovigatti,
  Doye, and Louis]{Sulc12}
\v{S}ulc, P.; Romano, F.; Ouldridge, T.E.; Rovigatti, L.; Doye, J.P.K.; Louis,
  A.A.
\newblock Introducing Sequence-Dependent Interactions into a Coarse-Grained
  {DNA} Model.
\newblock {\em J. Chem. Phys.} {\bf 2012}, {\em 137},~135101.

\bibitem[Snodin \em{et~al.}(2015)Snodin, Randisi, Mosayebi, Šulc, Schreck,
  Romano, Ouldridge, Tsukanov, Nir, Louis, and Doye]{Snodin15}
Snodin, B.E.K.; Randisi, F.; Mosayebi, M.; Šulc, P.; Schreck, J.S.; Romano,
  F.; Ouldridge, T.E.; Tsukanov, R.; Nir, E.; Louis, A.A.; Doye, J.P.K.
\newblock Introducing Improved Structural Properties and Salt Dependence into a
  Coarse-Grained Model of {DNA}.
\newblock {\em J. Chem. Phys.} {\bf 2015}, {\em 142},~234901.

\bibitem[Snodin \em{et~al.}(2019)Snodin, Schreck, Romano, Louis, and
  Doye]{Snodin19}
Snodin, B.E.K.; Schreck, J.S.; Romano, F.; Louis, A.A.; Doye, J.P.K.
\newblock Coarse-grained modelling of the structural properties of {DNA}
  origami.
\newblock {\em Nucleic Acids Res.} {\bf 2019}, {\em 47},~1585--1597.

\bibitem[Chhabra \em{et~al.}(2020)Chhabra, Mishra, Cao, Pre\v{s}ern, Skoruppa,
  Tortora, and Doye]{Chhabra20}
Chhabra, H.; Mishra, G.; Cao, Y.; Pre\v{s}ern, D.; Skoruppa, E.; Tortora,
  M.M.C.; Doye, J.P.K.
\newblock Computing the elastic mechanical properties of rod-like {DNA}
  nanostructures.
\newblock {\em J. Chem. Theory Comput.} {\bf 2020}, {\em 16},~7748--7763.

\bibitem[Benson \em{et~al.}(2018)Benson, Mohammed, Rayneau-Kirkhope, G{\aa}din,
  Orponen, and H\"{o}gberg]{Benson18}
Benson, E.; Mohammed, A.; Rayneau-Kirkhope, D.; G{\aa}din, A.; Orponen, P.;
  H\"{o}gberg, B.
\newblock Effects of design choices on the stiffness of wireframe {DNA} origami
  structures.
\newblock {\em ACS Nano} {\bf 2018}, {\em 12},~9291--9299.

\bibitem[Berengut \em{et~al.}(2019)Berengut, Berengut, Doye, Pre\v{s}ern,
  Kawamoto, Ruan, Wainwright, and Lee]{Berengut19}
Berengut, J.F.; Berengut, J.C.; Doye, J.P.K.; Pre\v{s}ern, D.; Kawamoto, A.;
  Ruan, J.; Wainwright, M.J.; Lee, L.K.
\newblock Design and synthesis of pleated {DNA} origami nanotubes with
  adjustable diameters.
\newblock {\em Nucleic Acids Res.} {\bf 2019}, {\em 47},~11963--11975.

\bibitem[Berengut \em{et~al.}(2020)Berengut, Wong, Berengut, Doye, Ouldridge,
  and Lee]{Berengut20}
Berengut, J.F.; Wong, C.; Berengut, J.C.; Doye, J.P.K.; Ouldridge, T.E.; Lee,
  L.K.
\newblock Self-limiting polymerization of {DNA} origami subunits with strain
  accumulation.
\newblock {\em ACS Nano} {\bf 2020}, {\em 14},~17428--17441.

\bibitem[Engel \em{et~al.}(2020)Engel, Romano, Louis, and Doye]{Engel20}
Engel, M.C.; Romano, F.; Louis, A.A.; Doye, J.P.K.
\newblock Measuring internal forces in single-stranded {DNA}: Application to a
  {DNA} force clamp.
\newblock {\em J. Chem. Theory Comput.} {\bf 2020}, {\em 16},~7764--7775.

\bibitem[Yao \em{et~al.}(2020)Yao, Zhang, Wang, Peng, Liu, Poppleton, \v{S}ulc,
  Jiang, Liu, Gong, Jing, Liu, Wang, Liu, Fan, and Yan]{Yao20}
Yao, G.; Zhang, F.; Wang, F.; Peng, T.; Liu, H.; Poppleton, E.; \v{S}ulc, P.;
  Jiang, S.; Liu, L.; Gong, C.; Jing, X.; Liu, X.; Wang, L.; Liu, Y.; Fan, C.;
  Yan, H.
\newblock Meta-DNA structures.
\newblock {\em Nat. Chem.} {\bf 2020}, {\em 12},~1067--1075.

\bibitem[Huang \em{et~al.}(2021)Huang, Kucinic, Johnson, Su, and
  Castro]{Huang21}
Huang, C.M.; Kucinic, A.; Johnson, J.A.; Su, H.J.; Castro, C.E.
\newblock Integrated computer-aided engineering and design for DNA assemblies.
\newblock {\em Nat. Mater.} {\bf 2021}, {\em 20},~1264--1271.

\bibitem[Li \em{et~al.}(2021)Li, Chen, Lee, and Choi]{Li21}
Li, R.; Chen, H.; Lee, H.; Choi, J.H.
\newblock Elucidating the mechanical energy for cyclization of a {DNA} origami
  tile.
\newblock {\em Appl. Sci.} {\bf 2021}, {\em 11},~2357.

\bibitem[Wong \em{et~al.}(arXiv:2108.06517)Wong, Tang, Schreck, and
  Doye]{Wong22}
Wong, C.K.; Tang, C.; Schreck, J.S.; Doye, J.P.K.
\newblock Characterizing the free-energy landscapes of {DNA} origamis, {
  arXiv:2108.06517}.

\bibitem[Kaufhold \em{et~al.}(arXiv:2110.01477)Kaufhold, Pfeifer, Castro, and
  Di~Michele]{Kaufhold22}
Kaufhold, W.T.; Pfeifer, W.; Castro, C.E.; Di~Michele, L.
\newblock Probing the mechanical properties of {DNA} nanostructures with
  metadynamics, { arXiv:2110.01477}.

\bibitem[Rovigatti \em{et~al.}(2015)Rovigatti, Šulc, Reguly, and
  Romano]{Rovigatti15}
Rovigatti, L.; Šulc, P.; Reguly, I.Z.; Romano, F.
\newblock A Comparison between Parallelization Approaches in Molecular Dynamics
  Simulations on GPUs.
\newblock {\em J. Comput. Chem.} {\bf 2015}, {\em 36},~1--8.

\bibitem[Kumar \em{et~al.}(1992)Kumar, Rosenberg, Bouzida, Swendsen, and
  Kollman]{Kumar92}
Kumar, S.; Rosenberg, J.M.; Bouzida, D.; Swendsen, R.H.; Kollman, P.A.
\newblock The Weighted Histogram Analysis Method for Free-Energy Calculations
  on Biomolecules. I. The Method.
\newblock {\em J. Comput. Chem.} {\bf 1992}, {\em 13},~1011--1021.

\bibitem[Marras \em{et~al.}(2016)Marras, Zhou, Kolliopoulos, Su, and
  Castro]{Marras16}
Marras, A.E.; Zhou, L.; Kolliopoulos, V.; Su, H.J.; Castro, C.E.
\newblock Directing folding pathways for multi-component {DNA} origami
  nanostructures with complex topology.
\newblock {\em New J. Phys} {\bf 2016}, {\em 18},~055005.

\bibitem[Wagenbauer \em{et~al.}(2014)Wagenbauer, Wachauf, and
  Dietz]{Wagenbauer14}
Wagenbauer, K.F.; Wachauf, C.H.; Dietz, H.
\newblock Quantifying quality in {DNA} self-assembly.
\newblock {\em Nat. Commun.} {\bf 2014}, {\em 5},~3691.

\bibitem[Myhrvold \em{et~al.}(2017)Myhrvold, Baym, Hanikel, Ong, Gootenberg,
  and Yin]{Myhrvold17}
Myhrvold, C.; Baym, M.; Hanikel, N.; Ong, L.L.; Gootenberg, J.S.; Yin, P.
\newblock Barcode extension for analysis and reconstruction of structures.
\newblock {\em Nat. Commun.} {\bf 2017}, {\em 8},~14698.

\bibitem[Strauss \em{et~al.}(2018)Strauss, Scheuder, Haas, Nickels, and
  Jungmann]{Strauss18}
Strauss, M.T.; Scheuder, F.; Haas, D.; Nickels, P.C.; Jungmann, R.
\newblock Quantifying absolute addressability in {DNA} origami with molecular
  resolution.
\newblock {\em Nat. Commun.} {\bf 2018}, {\em 9},~1600.

\end{thebibliography}


\end{document}


\title{Supplementary Material for ``The free-energy landscape of a mechanically bistable DNA origami''}
\author{Chak Kui Wong}
\affiliation{Physical and Theoretical Chemistry Laboratory, Department of Chemistry, University of Oxford, South Parks Road, Oxford, OX1 3QZ, United Kingdom}
\author{Jonathan P. K. Doye}
\affiliation{Physical and Theoretical Chemistry Laboratory, Department of Chemistry, University of Oxford, South Parks Road, Oxford, OX1 3QZ, United Kingdom}


\maketitle

\makeatletter
\renewcommand{\thefigure}{S\@arabic\c@figure}
\renewcommand{\theequation}{S\@arabic\c@equation}
\renewcommand{\thetable}{S\@arabic\c@table}
\renewcommand{\thesection}{S\@arabic\c@section}
 
\section{Further results}
\subsection{Free-energy contributions of each joint}
\begin{figure}[ht!]
    \centering
    \includegraphics[width=10.5cm]{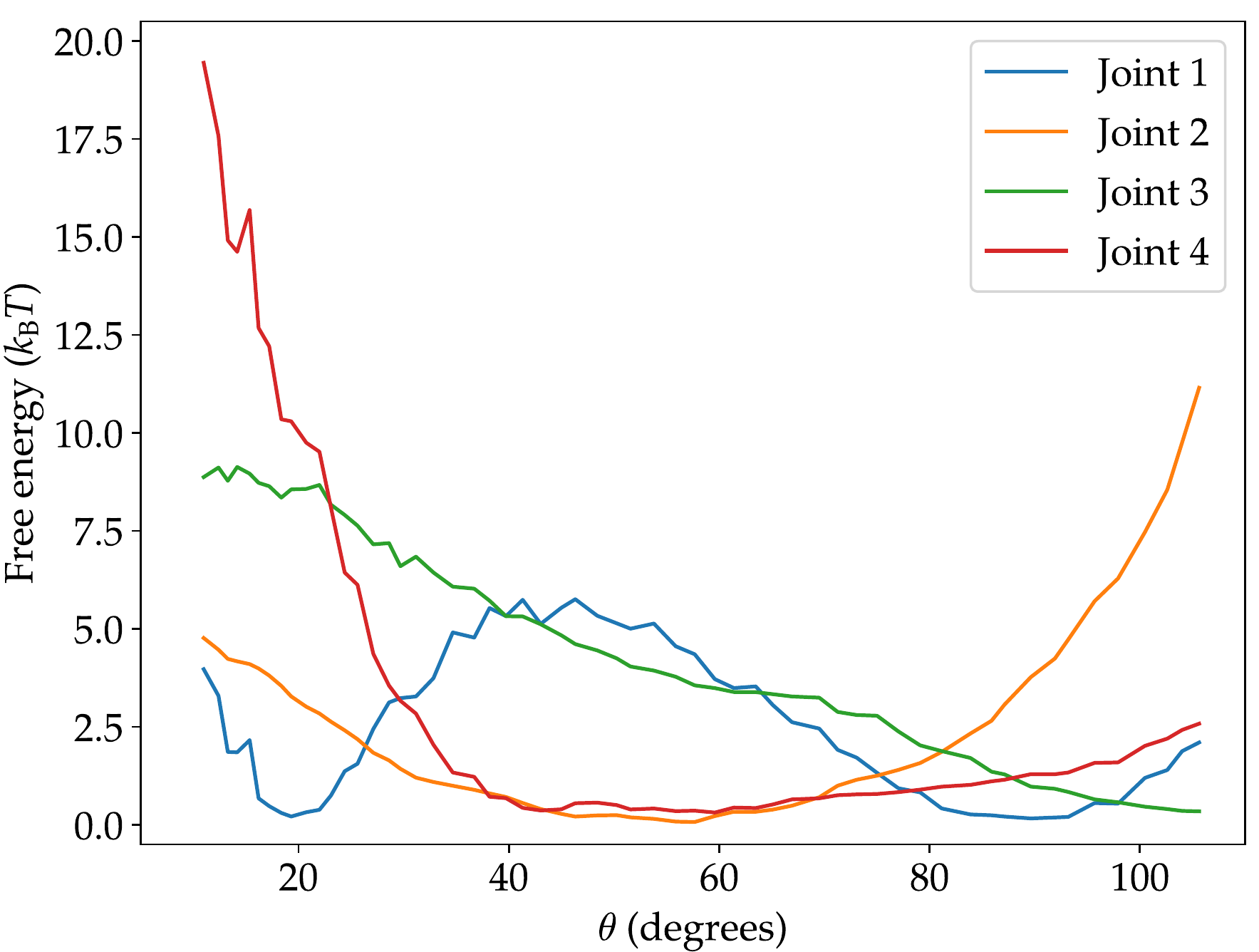}
    \caption{Free-energy contributions of each joint to that of the complete origami as a function of $\theta$.}
    \label{fig:castro_FEL_decomposed_each}
\end{figure}

\section{Simulation details}
We used the latest version of oxDNA (available at https://github.com/lorenzo-rovigatti/oxDNA) which supports an external harmonic potential that acts on the distance between the centres of mass of two groups of nucleotides.

Simulations were performed at 300\,K using a Langevin thermostat.
The time-step used was 0.005 in the internal simulation units of the oxDNA code, which corresponds to 15\,fs. 

\subsection{Generating starting configurations}
We converted the caDNAno design files of the origamis into oxDNA format using the tacoxDNA package.\cite{Suma19} The converted configurations are then rearranged manually in oxView \cite{Poppleton20} so that they are closer to their designed geometries.

The rearranged configurations cannot serve as starting configurations for molecular dynamics simulations because of nucleotides experiencing large forces due to particle overlaps or extended bonds. Therefore, the potential energy of these configurations is first minimized for 200 steps using a steepest-descent algorithm, and then the configurations are relaxed in a molecular dynamics simulation using a modified backbone potential for $10^6$ steps. After that, the extended bonds have typically returned to their normal lengths, and the configurations are ready for simulation using the standard oxDNA force field. All designs were equilibrated for a further $10^8$ steps, corresponding to about \SI{1.5}{\us}.

\subsection{Free-energy landscape}
\begin{figure}[ht!]
    \centering
    \includegraphics[width=\textwidth]{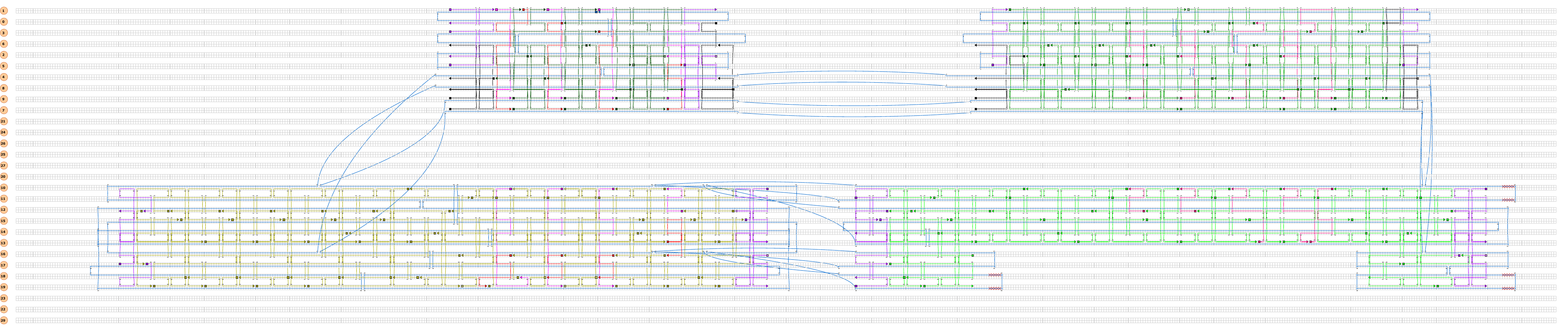}
    \caption{Full caDNAno design for the complete origami}
    \label{fig:all_cadnano}
\end{figure}

We used umbrella sampling \cite{Torrie77} to calculate the free-energy landscape as a function of the order parameter $R$, defined in Figure 2 of the main text.
Specifically, $R$ was defined as the distance between the centres of mass of two groups of nucleotides, which consist of the nucleotides at positions 259--273 on helices 0--9, and at positions 231--245 on helices 10--19, respectively. To obtain the landscape as a function of $\theta$, the values of $R$ and $\theta$ were tracked in each window of the umbrella sampling simulations. $\theta$ was defined as the angle between the two vectors in Figure 2 of the main text. 
Specifically, the first vector was defined as the vector pointing from the centre of mass of the nucleotides at positions 196--210 to that at positions 259--273 on helices 0--9, whereas the second vector was defined as the vector pointing from the centre of mass of the nucleotides at positions 147--161 to that at positions 231--245 on helices 10--19. The nucleotide positions and helix indices are defined in the caDNAno design in Figure \ref{fig:all_cadnano}.

We performed sets of simulations where the order parameter was restrained with a harmonic potential at the centre of each sampling window. The range of $R$ values sampled was from $R$ = \SI{12}{nm} to \SI{60}{nm}.

Before the umbrella sampling simulations, we prepared the starting configurations for each window using two pulling simulations. The order parameter of an equilibrated configuration was restrained at its original value (\SI{46}{nm}) with a harmonic bias potential of stiffness $k=$ \SI{57.09}{pN/nm}.

In the first simulation, the equilibrium position of the harmonic bias potential was gradually reduced at a constant rate of \SI{0.14}{m/s}, until it reached \SI{12}{nm}. Configurations were outputted every $2\times10^6$ steps, i.e.\ when the traps had moved by \SI{0.85}{nm}, resulting in a starting configuration for each window from the starting value of $R$ to the final value of $R$, where neighbouring windows are separated by \SI{0.85}{nm}. 

In the second simulation, the equilibrium position was gradually increased at the same rate as the first simulation, until it reached \SI{60}{nm}. Configurations were outputted at the same frequency, thus giving starting configurations for windows defined in a similar way as in the first simulation.

After the starting configurations have been prepared, the configuration in each window was restrained at the value of $R$ corresponding to that window with a harmonic bias potential of stiffness $k=\SI{11.42}{pN/nm}$. Each window was equilibrated for $10^6$ steps before a production run of $10^7$ steps. $R$ was outputted every $10^3$ steps, giving $10^4$ data points for each window. Using the biased probability distributions of $R$ in each window, we used WHAM \cite{Kumar92, wham} to calculate the unbiased free-energy landscape of the system as a function of $R$. 

A new set of production runs was then started from the last configurations of the previous production runs, and WHAM was performed on the new data points to calculate the free-energy landscape. This process was repeated three times. The resulting landscapes were very similar, thus confirming convergence.

\begin{figure}[ht!]
    \centering
    \includegraphics[width=10.5cm]{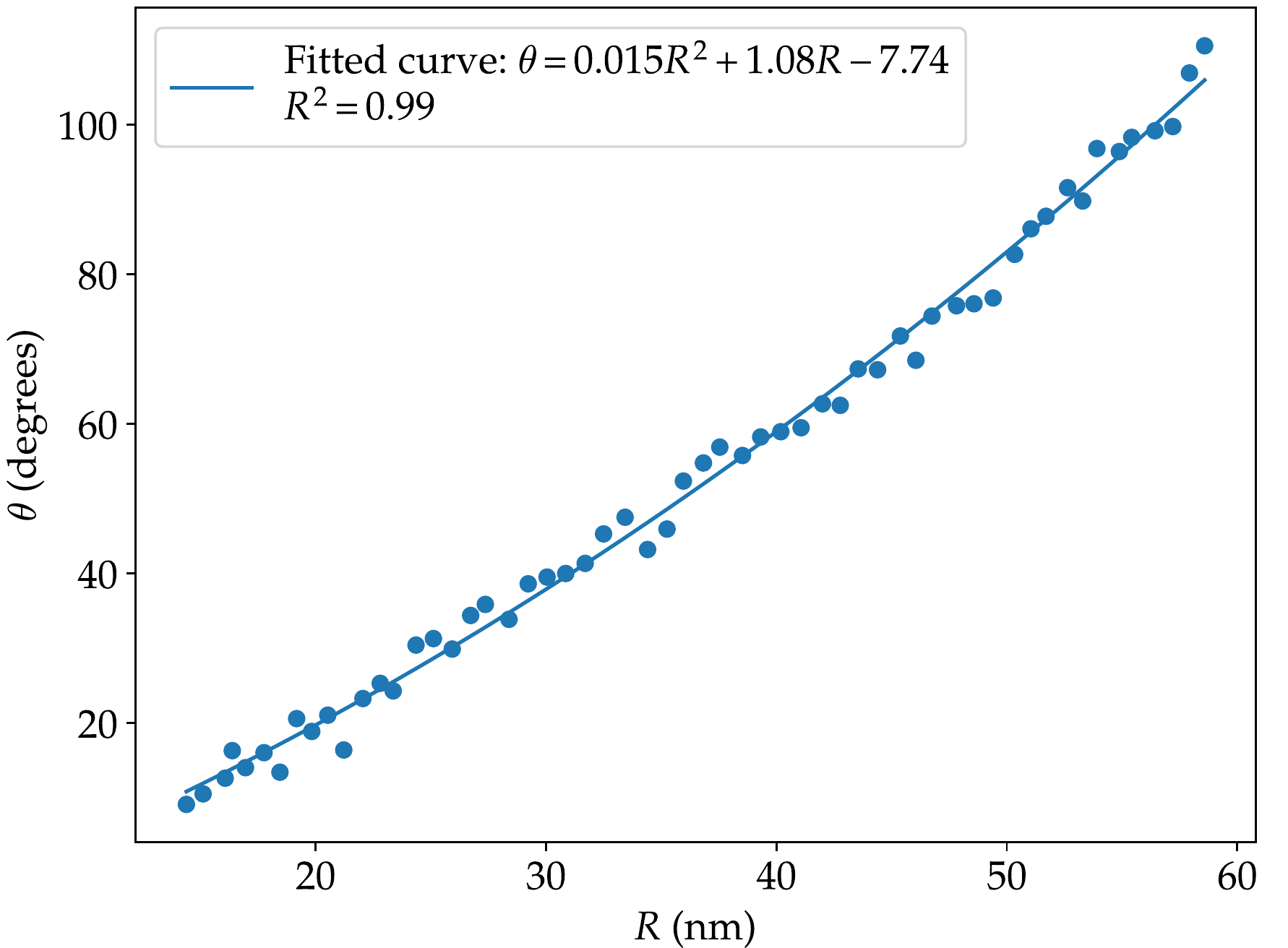}
    \caption{The mapping between the distance order parameter $R$ and $\theta$. The data points represent the average values of both for each umbrella sampling window and the line is a fit to this data.}
    \label{fig:castro_R_theta}
\end{figure}

In each umbrella sampling window, the value of $\theta$ was also recorded every $10^3$ steps. Using the corresponding values of $R$, a quadratic relationship between $R$ and $\theta$ was fitted and used to map the $x$-axis of the free-energy landscape from $R$ to $\theta$. The fitted relationship is $\theta = 0.015 R^2 + 1.08 R - 7.74$, with the coefficient of determination of the fit being 0.99. The data and the fit is shown in Fig.\ \ref{fig:castro_R_theta}.

\subsection{Free-energy decomposition}
We decomposed the free-energy landscape into four contributions associated with each of the four joints. 

\begin{figure}[ht!]
    \centering
    \includegraphics[width=\textwidth]{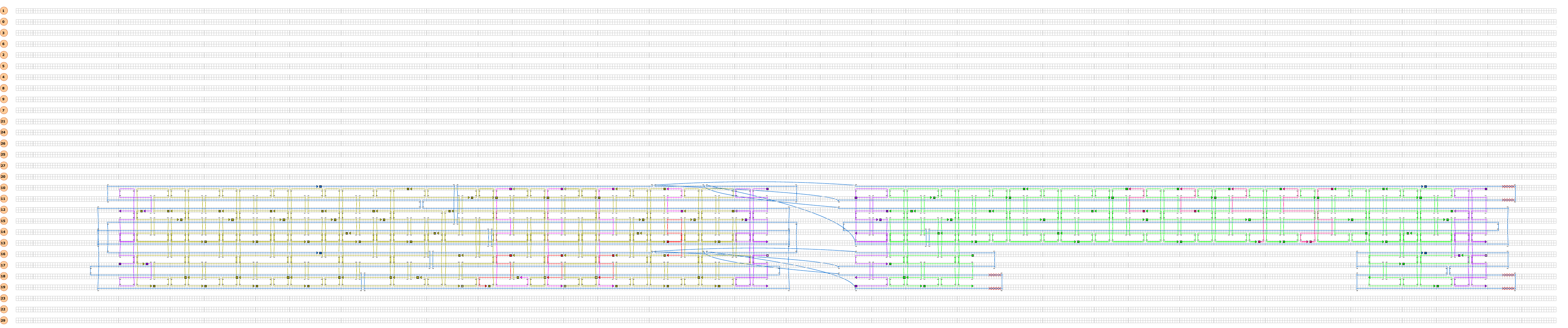}
    \caption{Full caDNAno design for the partial origami used for calculating the free-energy landscape of joint 1}
    \label{fig:none_cadnano}
\end{figure}

For joint 1, the origami design was modified from the original design by removing the nucleotides corresponding to the coupler and the crank blocks. The order parameter $R_1$ was defined as the distance between the centres of mass of two groups of nucleotides, which consist of the nucleotides at positions 147--161 on helices 10--19, and at positions 539--553 on helices 10--19, respectively. The nucleotide positions and helix indices are defined in the caDNAno design in Figure \ref{fig:none_cadnano}.

\begin{figure}[ht!]
    \centering
    \includegraphics[width=\textwidth]{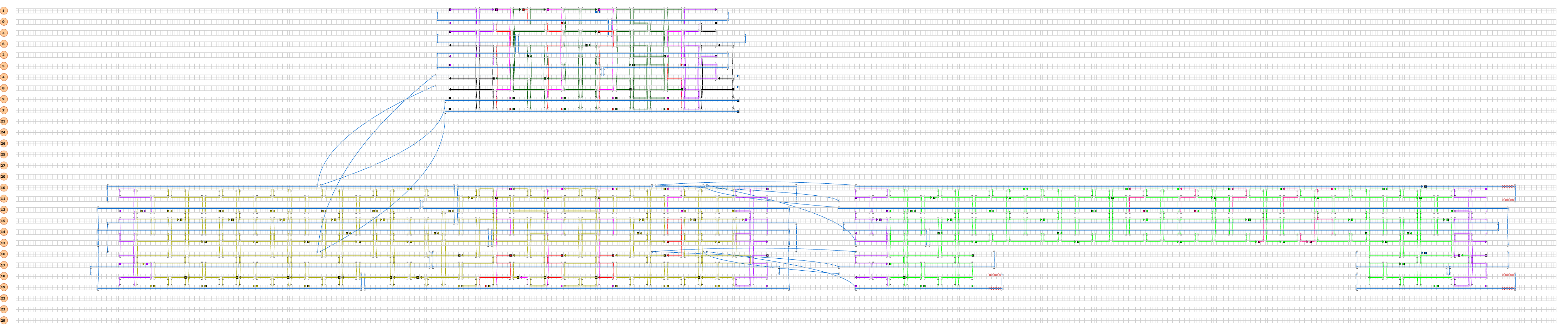}
    \caption{Full caDNAno design for the partial origami used for calculating the free-energy landscape of joint 2}
    \label{fig:lower_cadnano}
\end{figure}

For joint 2, the origami design was modified from the original design by removing the nucleotides corresponding to the coupler block. The order parameter $R_2$ was defined as the distance between the centres of mass of two groups of nucleotides, which consist of the nucleotides at positions 259--273 on helices 0--9, and at positions 231--245 on helices 10--19, respectively. The nucleotide positions and helix indices are defined in the caDNAno design in Figure \ref{fig:lower_cadnano}.

\begin{figure}[ht!]
    \centering
    \includegraphics[width=\textwidth]{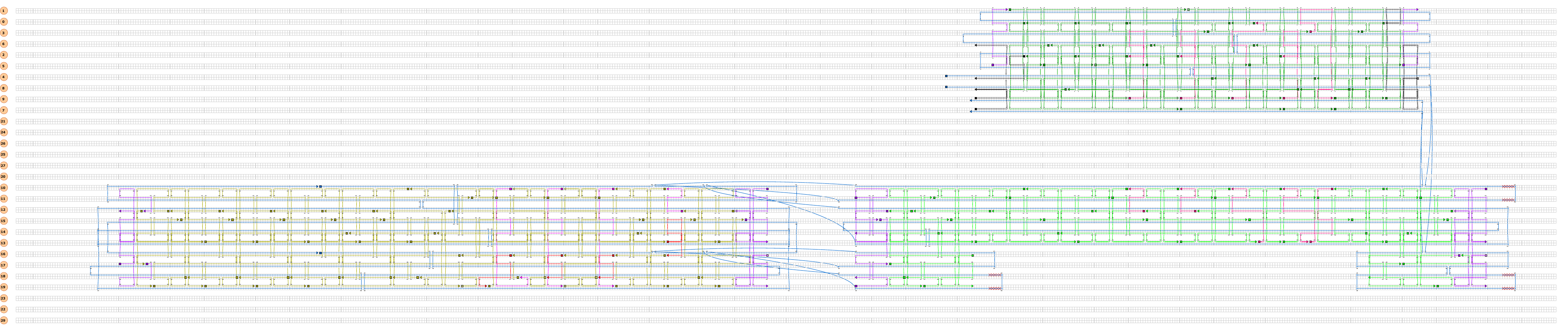}
    \caption{Full caDNAno design for the partial origami used for calculating the free-energy landscape of joint 3}
    \label{fig:upper_cadnano}
\end{figure}

For joint 3, the origami design was modified from the original design by removing the nucleotides corresponding to the crank block. The order parameter $R_3$ was defined as the distance between the centres of mass of two groups of nucleotides, which consist of the nucleotides at positions 413--427 on helices 0--9, and at positions 413--427 on helices 10--19, respectively. The nucleotide positions and helix indices are defined in the caDNAno design in Figure \ref{fig:upper_cadnano}.

\begin{figure}[ht!]
    \centering
    \includegraphics[width=\textwidth]{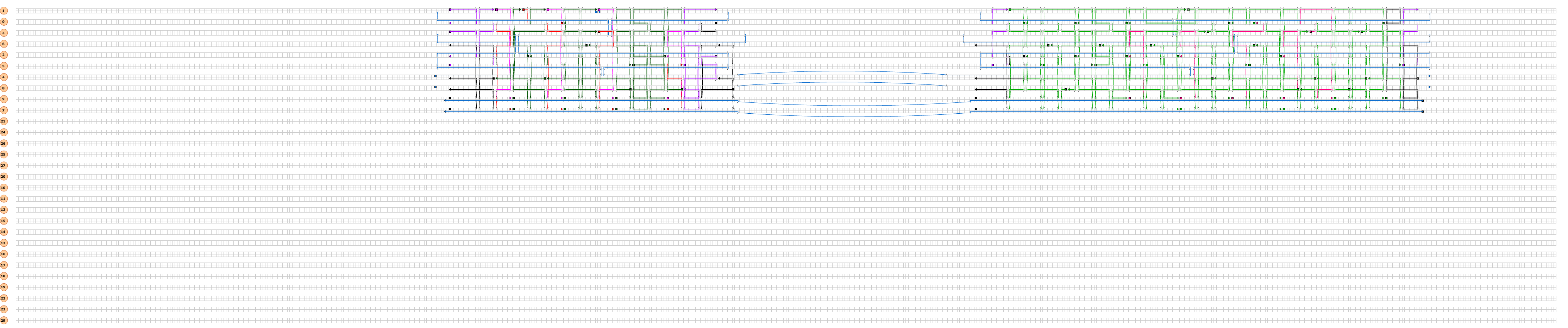}
    \caption{Full caDNAno design for the partial origami used for calculating the free-energy landscape of joint 4}
    \label{fig:both_cadnano}
\end{figure}

For joint 4, the origami design was modified from the original design by removing the nucleotides corresponding to the frame and compliant blocks. The order parameter $R_4$ was defined as the distance between the centres of mass of two groups of nucleotides, which consist of the nucleotides at positions 196--210 on helices 0--9, and at positions 539--553 on helices 0--9, respectively. The nucleotide positions and helix indices are defined in the caDNAno design in Figure \ref{fig:both_cadnano}.
Part 1 of the landscape corresponds to the part where the joint points outwards and resembles the origami in state S2, while part 2 corresponds to the part where the joint is folded inwards and resembles the origami in state S1. To calculate each sub-landscape we did not explicitly constrain the joint to point inwards or outwards. Instead, the portion of the landscape sampled was determined by the initial configurations, as for $R_4$ values corresponding to a bent joint, transitions between the two sub-landscapes are not feasible in the umbrella sampling simulations. For $R_4$ simulations corresponding to straight or stretched joints the two sets of simulations sample equivalent configurations.

For each joint, we performed a set of simulations where the order parameter was restrained with a harmonic potential in each sampling window. The sampling range of the order parameters was from 76--94\,nm for joint 1, 12--60\,nm for joint 2, 12--76\,nm for joint 3, 68--86\,nm for joint 4 (part 1), and 60--86\,nm for joint 4 (part 2). 

Umbrella sampling simulations were performed in a similar manner as those for the complete origami, with windows separated by \SI{0.85}{nm} and the configuration in each window being restrained with a harmonic potential of stiffness $k=$ \SI{11.42}{pN/nm}. Each window was equilibrated for $10^6$ steps before a production run of $10^7$ steps. $R_i$ was outputted every $10^3$ steps, giving $10^4$ data points for each window. Using the biased probability distributions of $R_i$ in each window, we used WHAM \cite{Kumar92, wham} to calculate the unbiased free-energy landscape of the system as a function of $R_i$. The production runs were repeated three times with the landscapes obtained being similar.

\begin{figure}[ht!]
    \centering
    \includegraphics[width=10.5cm]{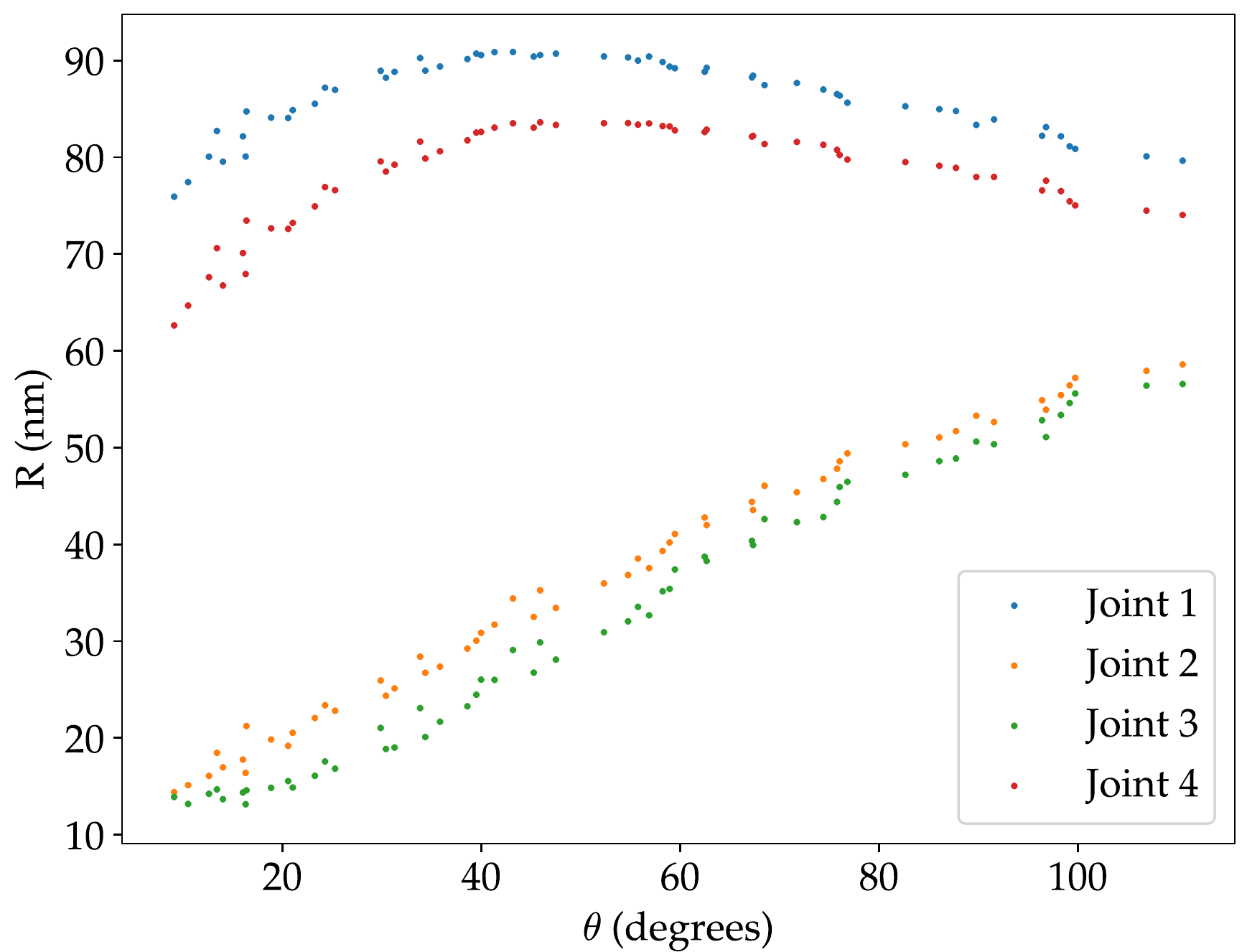}
    \caption{The mapping between the distance order parameters of the individual joints $R_i$ and $\theta$. The data points represent the average values for each umbrella sampling window of the complete origami.}
    \label{fig:castro_R_i_theta}
\end{figure}

To enable the transformation of the landscapes of the individual joints to be a function of $\theta$, the values of $R_i$, the distance order parameters for the joints, were also measured in the umbrella sampling simulations of the complete origami. Fig.\ \ref{fig:castro_R_i_theta} shows their variation with $\theta$. The maxima in $R_1$ and $R_4$ occur when the crank and coupler blocks are collinear. The two parts of the joint 4 landscape correspond to different ranges of $\theta$ with a small region of overlap corresponding to when the crank and coupler blocks are approximately collinear. To generate a single landscape as a function of $\theta$ for joint 4, part 1 was used for $\theta>53^\circ$ and part 2 for $\theta<53^\circ$ and the two landscapes matched at that point.

\clearpage
\section*{References}

%